\documentclass[aps,prx,twocolumn,superscriptaddress,showpacs]{revtex4-1}
\usepackage{graphicx}
\usepackage{bm}
\usepackage{color}
\usepackage{amssymb} 
\usepackage{amsmath}
\usepackage{verbatim}

\renewcommand\Re{\operatorname{Re}}
\renewcommand\Im{\operatorname{Im}}
\newcommand{\la}{\langle}
\newcommand{\ra}{\rangle}
\newcommand{\rt}{\right}
\newcommand{\lf}{\left}
\newcommand{\Tr}{\operatorname{Tr}}
\renewcommand{\k}{\mathbf k}
\newcommand{\s}{\text{s}}
\renewcommand{\i}{\text{in}}

\begin{document}
\title{Imaging instantaneous electron flow with ultrafast resonant x-ray scattering}
\author{Daria Popova-Gorelova}
\email[]{daria.gorelova@desy.de}
\affiliation{Center for Free-Electron Laser Science, DESY, Notkestrasse 85, D-22607 Hamburg, Germany}
\affiliation{The Hamburg Centre for Ultrafast Imaging, Luruper Chaussee 149, D-22761 Hamburg, Germany}
\author{Robin Santra}
\email[]{robin.santra@cfel.de}
\affiliation{Center for Free-Electron Laser Science, DESY, Notkestrasse 85, D-22607 Hamburg, Germany}
\affiliation{The Hamburg Centre for Ultrafast Imaging, Luruper Chaussee 149, D-22761 Hamburg, Germany}
\affiliation{Department of Physics, University of Hamburg, D-20355 Hamburg, Germany}
\date{\today}
\begin{abstract}
We propose a novel way to image dynamical properties of nonstationary electron systems using ultrafast resonant x-ray scattering. Employing a rigorous theoretical analysis within the framework of quantum electrodynamics, we demonstrate that a single scattering pattern from a nonstationary electron system encodes the instantaneous interatomic electron current in addition to the structural information usually obtained by resonant x-ray scattering from stationary systems. Thus, inelastic contributions that are indistinguishable from elastic processes induced by a broadband probe pulse, instead of being a concern, serve as an advantage  for time-resolved resonant x-ray scattering. Thereby, we propose an approach combining elastic and inelastic resonant x-ray scattering for imaging dynamics of nonstationary electron systems in both real space and real time. In order to illustrate its power, we show how it can be applied to image the electron hole current in an ionized diatomic molecule.
\end{abstract}
\maketitle

\section{Introduction}

Electron dynamics in valence shells of atoms and molecules determine various physical processes, such as chemical reactions, cooperative phenomena in solids, charge migration in biological systems {\it etc}. Imaging of electron dynamics both in real time and real space is one of the most important goals for modern ultrafast science \cite{KrauszRMP09, SmirnovaNature09, GoulielmakisNature10, SansoneNature10, HaesslerNature10, TzallasNature11, HockettNature11, CooperPRL13, LeeuwenburghPRL13}. X-ray free-electron lasers are a promising tool to achieve this task \cite{CorkumNature07, GaffneyScience07, ChapmanNature06, VrakkingNature12, LeoneNature14}. Wavelengths of hard x rays provide angstrom resolution, which corresponds to inter-atomic distances in molecular structures and solids. At the same time, free-electron laser sources are able to produce ultrashort high intensity pulses, which give access to femtosecond time scales \cite{EmmaNature10, McNeilNature10}. 

While technological developments approach the possibility to make `electron movies', it is necessary to answer fundamental questions about the interaction between a nonstationary electron system and an ultrashort light pulse.  In this paper, we analyze time-resolved diffraction imaging by resonant hard x-ray scattering from nonstationary electron systems and demonstrate that the instantaneous interatomic electron current is encoded in the Fourier transform of a scattering pattern. Resonant x-ray scattering (RXS) is a powerful technique that provides insight into charge, orbital and spin degrees of freedom \cite{FinkRPP13,MatsumuraJPSJ13,DmitrienkoActaCrysA05,LoveseyPhRep05}. RXS is an element specific method, since it involves transitions from atomic core shells with resonant excitation energies that strongly depend on the atomic species. The resonant nature of this process allows one to considerably enhance the scattering cross section in comparison to the nonresonant case studied in Refs.~\cite{DixitPNAS12, ZamponiPNAS12}. This is particularly relevant for the measurement of valence electron dynamics in heavy elements, where the vast majority of electrons are stationary, since nonresonant x-ray scattering probes simultaneously electrons involved in the dynamics and electrons that are essentially stationary \cite{DixitJChPh13}. In resonant scattering, one can selectively enhance the scattering contribution from those (quasi-)particles that are actually moving. Resonant conditions are also essential for magnetic scattering, since magnetic interactions with light are very weak \cite{HannonPRL88}. Time-resolved diffraction by resonant x-ray pulses has been used to reveal various ultrafast phenomena, such as melting of orbital and spin orders in strongly correlated materials \cite{PontiusAPL11, EhrkePRL11}, laser-induced spin reversal \cite{GravesNature13} and demagnetization \cite{VodungboNature12}. Applications to coherent electron dynamics have not yet been reported.

X-ray Raman scattering has been proposed as a spectroscopic probe of valence excited states in molecules \cite{TanakaPRL02,SchweigertPRA07}. It has been suggested that information about valence electron dynamics can be obtained by analyzing Raman spectroscopy signals, such as the change in transmission of a probe pulse with respect to a pump pulse \cite{MukamelARPC13}. We propose a different method for measuring electron dynamics, which is momentum- and time-resolved resonant diffraction by hard x rays. The advantage of this method is that one can directly image in both real time and real space electron dynamics such as charge migration, electronic wave packets in molecules, charge transfer during chemical reactions, as well as collective electron excitations.

The differential scattering probability (DSP) of RXS from a stationary crystal is dominated by elastic scattering processes, since their amplitudes sum up coherently giving rise to charge or magnetic Bragg peaks. Therefore, a most straightforward approach to calculate the DSP from an electron wave packet is to consider only the contribution of the elastic scattering processes, which do not change the state of the wave packet.  However, it has recently been demonstrated that inelastic (Compton-type) processes considerably affect scattering patterns obtained by ultrafast nonresonant x-ray scattering \cite{DixitPNAS12}  and electron diffraction \cite{ShaoPRA13}, leading to a loss of information about the instantaneous electron density. Thus, we develop a description of RXS from nonstationary electron systems based on quantum electrodynamics (QED), since it allows taking into account both elastic and inelastic processes correctly \cite{TanakaPRA01, HenriksenJPhCh08}.  Although the role of inelastic processes for time-resolved {\it nonresonant} x-ray scattering has been already analyzed \cite{DixitPNAS12, DixitJChPh13, DixitPRL13, DixitPRA14}, these studies cannot be applied to the {\it resonant} case, since high-energy resonant and nonresonant x-ray scattering are determined by different terms of the light-matter interaction Hamiltonian. Moreover, we find that in contrast to nonresonant x-ray scattering \cite{DixitPNAS12}, time-resolved RXS allows resolving the direction of electron flow without losing the connection to structural information despite inelastic contributions. Thereby, we show that although inelastic scattering processes are usually used for a spectroscopy analysis, and only pure elastic RXS is used to image structural information in stationary measurements, inelastic contributions serve as an advantage for  time-resolved RXS and provide additional insights into electron dynamics.


In the next Section, we derive the QED description of the DSP of time-resolved RXS that takes into account both elastic and inelastic processes and compare it to a ``quasi-stationary" description that assumes that the contribution from elastic processes to a time-resolved scattering pattern dominates. 
In Section \ref{section_probcurr}, we show the connection between the Fourier transform of a scattering pattern and the interatomic electron current at the time of measurement. We illustrate our results by describing a possible experiment in Section \ref{DiatMolecule}, showing how the instantaneous electron current in ionized Br$_2$ molecule can be imaged.

\section{Differential scattering probability from a nonstationary electron system}

Let us consider an electron system with Hamiltonian $\hat H_{\text{m}}$, which is the many-body electronic Hamiltonian in the absence of an x-ray field, with eigenstates $|\Phi_I\ra$ and eigenenergies $E_{I}$. We investigate scattering patterns of a resonant x-ray probe pulse from an electron system that had been excited by a pump pulse into a coherent superposition of the electronic eigenstates at time $t=0$. The time evolution of the coherent superposition is given by
\begin{equation}
|\Psi(t) \ra = \sum_I C_Ie^{-iE_It} |\Phi_I\ra.
\end{equation}
We assume that the pump and probe pulses do not overlap in time. In this way, it becomes possible to describe the probe and the pump steps separately (see \cite{SantraPRA11}). Since the goal of this paper is to describe how one can measure electron dynamics of a given nonstationary electron system, the specific pump process giving rise to $|\Psi(t) \ra$ is of no concern here. In Section \ref{Section_General}, the description of scattering patterns from a coherent wave packet will be generalized for a statistical mixture of states.

\subsection{Quasi-stationary description}

The semiclassical approach to calculate the DSP of RXS from a wave packet  is to substitute  the wave packet state at the time of measurement $t_p$, $|\Psi(t_p)\ra=\sum_IC_Ie^{-iE_I t_p}|\Phi_I\ra$, for the ground state in the general relation for stationary RXS \cite{MaPRB94}. Then,
\begin{align}
&{\frac{dP}{d\Omega}}^{\text{(st)}} =\frac{\int dt I_{\i}(t)}{\omega_\i}\sum_{s_\s=1}^2\frac{\omega_\i^4}{c^4}\lf|\sum_{C}f_C(t_p)e^{i\mathbf Q\cdot \mathbf R_C} \rt|^2,\label{DSP_st}\\
&f_C(t_p)=\sum_{J_C}\frac{ \la \Psi(t_p)|\boldsymbol\epsilon_s^*\cdot\mathbf r|\Phi_{J_C}\ra\la\Phi_{J_C}|\boldsymbol\epsilon_\i\cdot\mathbf r| \Psi(t_p)\ra}
{(\omega_\i-(E_{J_C}-\la E \ra))+i\Gamma_{J_C}/2},\nonumber
\end{align}
where $\mathbf Q$ is the scattering vector, $c$ is the speed of light, $I_{\i}(t)$ is the probe pulse intensity, $\omega_\i$ is the photon energy of the incoming beam, $\la E \ra$ is the mean energy of the electron wave packet and $\boldsymbol\epsilon_\i$ is the mean polarization vector of the incoming beam (we use atomic units for this and the following expressions). The sum over $s_\s$ denotes the sum over the polarization vectors $\boldsymbol\epsilon_s$ of the scattered photons. In Eq.~(\ref{DSP_st}), $f_C(t_p)$ is the scattering amplitude of atom $C$ situated at position $\mathbf R_C$. $J_C$ denotes an intermediate state with a hole in a core shell of atom $C$, $\Gamma_{J_C}$ is the decay width of $J_C$. The spatial distribution of the x-ray electric field is treated within the dipole approximation for each absorbing atom, since we consider a resonant process involving transitions of electrons from core shells, which are highly localized in comparison to the resonant wavelengths considered here.

\subsection{QED description}

Let us compare the result obtained by the quasistationary description in Eq.~(\ref{DSP_st}) to the DSP derived from QED. If the electron system is probed by an x-ray pulse, the total Hamiltonian of the whole system, matter and light, can be written as \cite{Loudon}
\begin{equation}
\hat H = \hat H_{\text{m}}+\sum_{\mathbf{k},s}\omega_{\k,s}\hat a_{\k,s}^\dagger\hat a_{\k,s}+\hat H_{\text{int}},
\end{equation}
where $\hat a_{\k,s}^\dagger$ and $\hat a_{\k,s}$ are creation and annihilation operators of a photon in the $\k$, $s$ mode of the radiation field with energy $\omega_{\k}=|\k|c$. 
$\hat H_{\text{int}}$ is the minimal coupling interaction Hamiltonian between the matter and the electromagnetic field in Coulomb gauge
\begin{equation}
\hat H_{\text{int}}=
\frac1c\int d^3r\hat \psi^\dagger(\mathbf r)\lf(\hat{\mathbf A}(\mathbf r)\cdot\mathbf p\rt)\hat \psi(\mathbf r),\label{H_int}
\end{equation}
where $\hat{\mathbf A}$ is the vector potential of the electromagnetic field, $\mathbf p$ is the canonical momentum of an electron, $\hat \psi^\dagger$ and $\hat \psi$ are electron creation and annihilation field operators. We ignore the term of the interaction Hamiltonian determined by $\hat{\mathbf A}^2$ (which dominates for high-energy nonresonant x-ray scattering), since it becomes negligible compared to Eq.~(\ref{H_int}) in the case of resonant scattering. 
The DSP is connected to the probability $P(\k_{\s})$ of observing a scattered photon with momentum $\k_{\s}$, which differs from the incoming photon momenta, by
\begin{equation}
\frac{dP}{d\Omega} = \frac{V}{(2\pi c)^3}\int_{0}^{\infty}d\omega_{\k_\s}\omega_{\k_\s}^2P(\k_{\s})\label{DSP},
\end{equation}
where $V$ is the quantization volume.  $P(\k_{\s})$ is given within the density matrix formalism \cite{Mandel} by
\begin{equation}
P(\mathbf k_s)=\lim_{t_f\to+\infty}\Tr\lf [\hat O_{\mathbf k_s}\hat\rho_f(t_f)\rt]\label{Probab},
\end{equation}
where the operator
\begin{equation}
\hat O_{\mathbf k_s} =\sum_{s_s=1}^2W(\omega_{\k_\s}) \sum_F\sum_{\{n'\}}|\Phi_F;\{n'\}\ra\la\Phi_F;\{n'\}|\label{Oks}
\end{equation}
describes the observation of a photon in the scattering mode $\k_\s$, $W(\omega_{\k_\s})$ represents the spectral acceptance range of the  photon detector, $\{n'\}$ is a field configuration that has one photon in the scattering mode $\k_{\s}$, $|\Phi_F\ra$ is a final electronic state vector. $\hat\rho_f(t_f)$ is the total density matrix of the electron system and the electromagnetic field at time $t_f$ after the action of the probe pulse, which we evaluate within the second-order time-dependent perturbation theory using $\hat H_{\text{int}}$ as the perturbation. We take into account that the probe pulse duration, $\tau_p$, must be much shorter than the time variation of the wave packet during the action of the probe pulse (see Eq.~(\ref{UltrashortPulseApr_App})). That means that the spectral bandwidth of the probe pulse must be much larger than the maximum energy splitting among the electron states involved in the dynamics. Therefore, it becomes impossible to separate elastic and inelastic scattering events through the spectroscopy of the scattered photon.

We show in Appendix \ref{Appendix_DSP} that, with those approximations, the DSP of a probe pulse with intensity $I_\i(t) = I_0\,e^{-4\ln2 ((t-t_p)/\tau_p)^2}$, which arrives at time $t_p$ after the pump pulse  can be written as
\begin{align}
&\frac{dP}{d\Omega}= \frac{ I_0\tau_p^2}{4\ln2c^4}\int_0^{\infty}d\omega_{\k_\s}\omega_{\k_\s}W(\omega_{\k_\s})\sum_{F,s_\s}\Biggl|\sum_{J_C}\Delta\omega_{J_CF}
\label{DSP_DipApp}\\
&\times\frac{\la \Phi_F|\boldsymbol\epsilon_\s^*\cdot\mathbf r|\Phi_{J_C}\ra\la\Phi_{J_C}|\boldsymbol\epsilon_\i\cdot\mathbf r|\Psi(t_p)\ra }{(\omega_{\k_\s}-\Delta\omega_{J_CF})+i\Gamma_{J_C}/2} e^{i\mathbf Q\cdot\mathbf R_{C}}\Biggr|^2 e^{-\frac{\Omega_F^2\tau_p^2}{4\ln2}},
\nonumber
\end{align}
where $\Delta\omega_{J_CF}=E_{J_C}-E_F$ and $\Omega_F = \omega_{\k_\s}-\omega_{\i}+E_F-\la E \ra$. 
Equation~(\ref{DSP_DipApp}) as well as Eq.~(\ref{DSP_st}) describe a process where the system is brought into some intermediate state $J_C$ by absorption of a photon, and then to some final state by spontaneous emission. However, the key difference between the expressions is that the wave-packet state $|\Psi(t_p)\ra$ enters Eq.~(\ref{DSP_DipApp}) only once.  
The quasi-stationary description in Eq.~(\ref{DSP_st}) applies an analogy to an elastic process in a stationary system and assumes that the system is brought back into the same electronic wave-packet state after the scattering process. However, this analogy is not correct, since the initial state of a nonstationary system is a superposition of its eigenstates, and the elastic process resulting in the final state that is exactly the same superposition of states cannot be spectroscopically distinguished.  This can also be shown from another perspective. The absorption process, bringing the system into some intermediate state, destroys the wave packet. Thus, the information about the wave packet is contained only in the absorption term $\la\Phi_{J_C}|\boldsymbol\epsilon_\i\cdot\mathbf r|\Psi(t_p)\ra$, but not in the spontaneous emission term, $\la \Phi_F|\boldsymbol\epsilon_\s^*\cdot\mathbf r|\Phi_{J_C}\ra$. However, the emission step is not possible, if absorption has not taken place. Thereby, the emission indirectly depends on the wave packet and a scattered photon indeed provides the information about the nonstationary electronic system.

It is shown in Appendix \ref{App_Stationary} that Equation~(\ref{DSP_DipApp}) applied to a stationary system at $\tau_p\gg1/\Gamma_{J_C}$ and $W(\omega_{\k_\s})=1$ goes over into the conventional relation for RXS, which takes into account elastic and inelastic contributions \cite{MaPRB94}. Although Eq.~(\ref{DSP_DipApp}) may resemble the expression for stationary inelastic RXS, there are fundamental differences between them. In a time-resolved measurement, where the pulse duration must be much shorter than the characteristic time of the electron dynamics, the incoming photon energy is  not precisely defined, and the signal depends on the spectrum of the broadband probe pulse \cite{DixitPRA14}. Spectrally resolved inelastic x-ray scattering from a stationary system encodes a dynamic structure factor, which can provide access to electron dynamics  at the level of linear response theory \cite{AbbamontePRL04, ReedScience10}. Our method measures the state of an electron system at a given time independently from how this state had been created. Another difference is that, in contrast to the inelastic x-ray scattering technique \cite{AmentRMP11}, energy resolution is not required in our method in order to obtain information about electron dynamics, as will be shown in Section \ref{section_probcurr}.

\subsection{General expression}

\label{Section_General}
If the state prepared by the pump pulse is a statistical mixture rather than a coherent wave packet, the system at time $t$ must be described by the density matrix
\begin{equation}
\hat \rho^{\text{m}}(t) = \sum_{I,K} \mathcal I_{IK}(t) |\Phi_I\ra\la \Phi_K| \label{El_den}.
\end{equation} 
In this case, the expression for the DSP becomes
\begin{align}
\frac{dP}{d\Omega}= &\frac{\tau_p^2I_0}{4\ln2c^4}\sum_{C_q,C_r}e^{i\mathbf Q\cdot(\mathbf R_{C_q}-\mathbf R_{C_r})}\nonumber\\
&\times\sum_{I,K}\mathcal I_{IK}(t_p)\la \Phi_K |  \hat {\mathcal G}_{qr} |\Phi_I\ra \label{DSP_DipApp_extended},
\end{align}
where $\hat {\mathcal G}_{qr}$ is provided in Eq.~(\ref{Gqr_App}). This expression generalizes Eq.~(\ref{DSP_DipApp}) and reduces to Eq.~(\ref{DSP_DipApp}) when  $\mathcal I_{IK}(t) = C_IC_K^*e^{-i(E_I-E_K)t}$, which is the condition that $\hat \rho^{\text{m}}(t)$ describes a perfectly coherent wave packet. 

\section{Connection to the probability current density}
\label{section_probcurr}

In contrast to the quasi-stationary description in Eq.~(\ref{DSP_st}), which straightforwardly generalizes the elastic x-ray scattering theory, the DSP from a nonstationary system in Eq.~(\ref{DSP_DipApp}) is not determined by the structure factor at the time of measurement, $\sum_Cf_C(t_p)e^{i\mathbf Q\cdot\mathbf R_C}$. Thus, let us consider what information is actually encoded in such a scattering pattern. First, structural information is still present due to the factors $e^{i\mathbf Q\cdot\mathbf R_C}$. Second, as will be shown below, the probability current between scattering atoms at the time of measurement is encoded in a scattering pattern. 

For the general situation in Eq.~(\ref{El_den}), the probability current density at the time of measurement, given by
\begin{align}
\mathbf j(\mathbf r, t_p) =\frac{i}{2}\text{Tr}\lf\{\hat \rho^{\text{m}}(t_p)\lf([\boldsymbol\nabla\hat\psi^\dagger]\hat\psi-\hat\psi^\dagger[\boldsymbol\nabla\hat\psi]\rt) \rt\} \label{ProbCurrDen}
\end{align}
can be decomposed into intra-atomic and inter-atomic contributions as
\begin{align}
&\mathbf j(\mathbf r, t_p) = \sum_{C_a} \mathbf j_{a}^{\text{intra}}(\mathbf r, t_p) + \sum_{C_a,C_b\neq C_a} \mathbf j_{ab}^{\text{inter}}(\mathbf r, t_p),
\end{align}
where $C_a$ and $C_b$ refer to all atoms in the system (see Appendix~\ref{App_Current}). The volume-integrated probability current between scattering atoms $C_q$ and $C_r$ is given by
\begin{align}
j_{qr}(t_p) &= \int d^3 r \, \mathbf j_{qr}^{\text{inter}}(\mathbf r, t_p)\cdot\mathbf n_{qr} \label{InterAtCurr}\\
 &= \Im\lf(\sum_{I,K}\mathcal I_{IK}(t_p)\la\Phi_K|\hat G_{qr}|\Phi_I\ra\rt)\nonumber,
\end{align} 
where  $\mathbf n_{qr}$ is the unit vector pointing from site $C_q$ to site $C_r$, and $\hat G_{qr} =  \int d^3 r \hat\xi_{C_r}^\dagger(\mathbf r)(\boldsymbol\nabla\cdot\mathbf n_{qr})\hat\xi_{C_q}(\mathbf r)$, the operator $\hat\xi_{C}$ annihilating a particle at site $\mathbf R_C$. 

We find that the interatomic probability current is encoded in the Fourier transform of the DSP, 
\begin{align}
\mathcal F_{\text{DSP}}(\mathbf r)=\frac{1}{(2\pi)^3}\int d^3 Q  \frac{dP(\mathbf Q)}{d\Omega}e^{-i\mathbf Q\cdot \mathbf r}.
\end{align}
Namely, the imaginary part of the Fourier transform of the general expression (\ref{DSP_DipApp_extended}) for the DSP,
\begin{align}
\Im(\mathcal F_{\text{DSP}}(\mathbf r)) \propto \sum_{C_q,C_r}\mathcal J_{qr} (t_p)\delta[\mathbf r -( \mathbf R_{C_q}-\mathbf R_{C_r})],
\end{align}
has delta peaks at the interatomic distances between scattering atoms $C_q$ and $C_r$ weighted by the factor
\begin{align}
\mathcal J_{qr}(t_p)
&= \Im\lf(\sum_{I,K}\mathcal I_{IK}(t_p)\la \Phi_K|\mathcal {\hat G}_{qr}|\Phi_I\ra\rt)\label{J_qr},
\end{align}
where $\mathcal {\hat G}_{qr} = \hat d_{C_r}^\dagger \hat S  \hat d_{C_q}$ [see Eq.~(\ref{Gqr_App}) for $\hat d_{C_{q,r}}$ and $\hat S$]. 
The time-dependent factors $\mathcal I_{IK}(t_p)$ entering in $\mathcal J_{qr}(t_p)$ and $j_{qr}(t_p)$ are weighted in both cases by matrix elements that are determined by the amounts of charges at sites $C_q$ and $C_r$. Namely,  the matrix element $\la \Phi_K|\hat G_{qr}|\Phi_I\ra$ depends on the amount of charges at sites $C_{q}$ and $C_r$ due to the operators $\hat\xi_{C_{q}}$ and  $\hat\xi_{C_{r}}$ that single out these charges. The operators $\hat d_{C_{q}}$ and $\hat d_{C_{r}}$ in $\la \Phi_K|\mathcal {\hat G}_{qr}|\Phi_I\ra$ select the charges via dipole matrix elements of transitions to and from inner shells localized at sites $C_{q}$ and $C_r$ that are determined by the charges at these sites. The operator $\hat S$ gives a time-independent contribution that can be factored out from the sum over $I$ and $K$ and, consequently, does not influence the temporal behavior of $\mathcal J_{qr}(t_p)$. As a result, the temporal evolution of the factor $\mathcal J_{qr}(t_p)$ substantially reproduces the temporal behavior of the interatomic current $j_{qr}(t_p)$ (see the discussion in Appendix \ref{Connection_SecApp} for additional details). This relation between the factor $\mathcal J_{qr}$ and the current $j_{qr}$ is analogous to the relation between the scattering amplitude $f_C$ and the charge at site $C$ in stationary elastic RXS. In the next Section, we provide an example showing the connection between $\mathcal J_{qr}(t_p)$ and $j_{qr}(t_p)$ in an ionized diatomic molecule.

\section{Application to a diatomic homonuclear molecule}
\label{DiatMolecule}
  
\begin{figure}[t]
\includegraphics[width=0.4\textwidth]{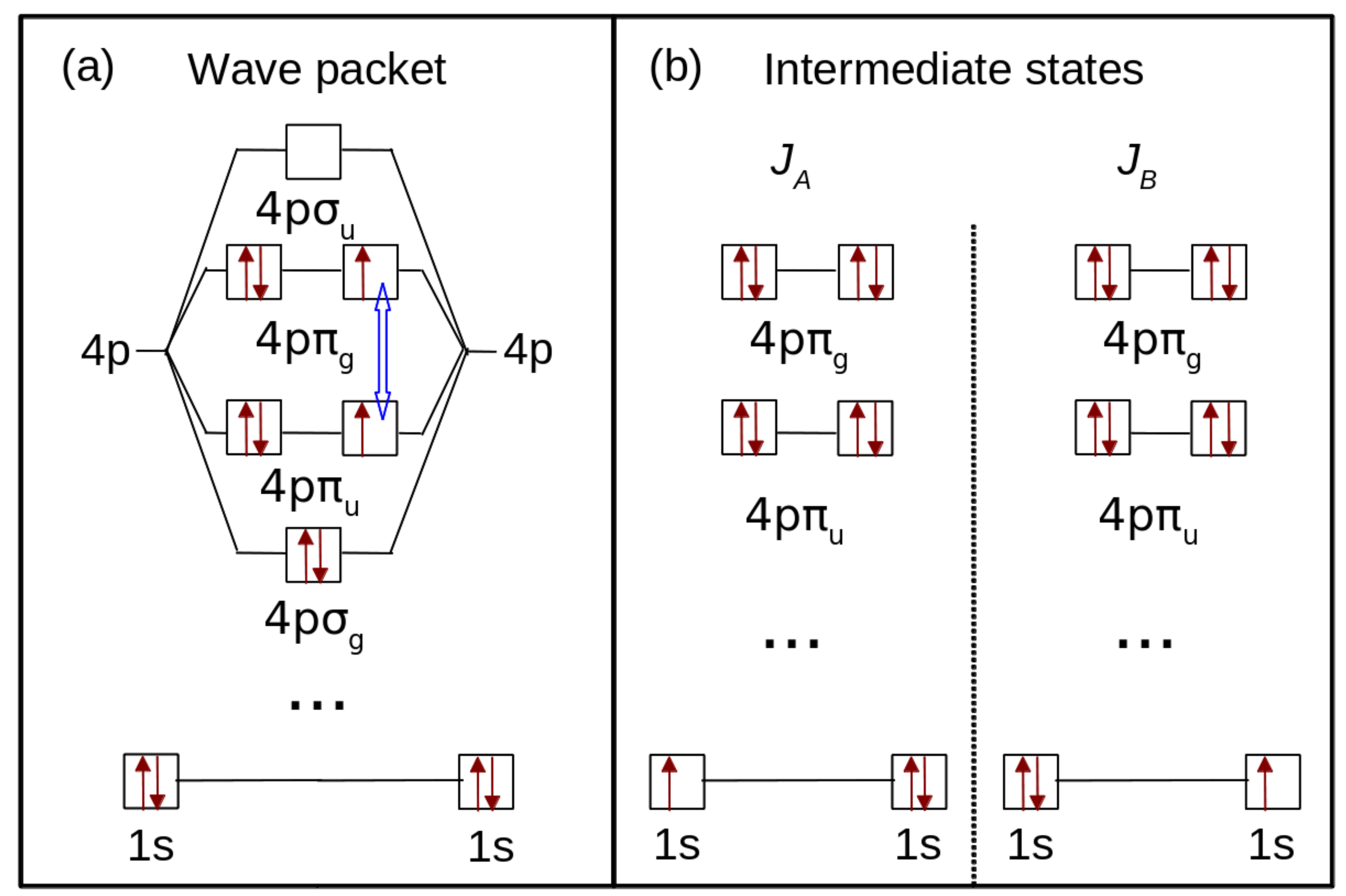}
  \caption{Molecular orbital diagram of the relevant states of Br$_2^+$. (a) The wave packet: an electron hole in the superposition of the $4p\pi_g$ and $4p\pi_u$ orbitals. (b) The intermediate states $J_A$ and $J_B$: the closed valence shell and an electron hole in either of the two Br $1s$ orbitals.}
\label{Fig_Br}
\end{figure}

In order to demonstrate the basic properties of the time-resolved scattering patterns and compare the results of Eqs.~(\ref{DSP_st}) and (\ref{DSP_DipApp}), we apply our formalism to RXS from an electronic wave packet in a diatomic homonuclear molecule. In recent years,  significant progress in attosecond science has made it possible to prepare in a controlled way electronic wave packets in molecules with characteristic time scales of femtoseconds to sub-femtoseconds. It is possible to launch a wave packet by removing one or two valence electrons from desired orbitals \cite{SmirnovaNature09,GoulielmakisNature10,SansoneNature10, HaesslerNature10,TzallasNature11} and even control the outcome of a simple chemical reaction \cite{RanitovicPNAS14} with ultrashort pump pulses. Strong-field ionization with subcycle optical field transients provides attosecond temporal confinement of ionization and enables a consequent triggering of a wave packet with a well-defined phase \cite{WirthScience11, ChiniNature14}. It has been demonstrated that valence-electron wave packets can evolve with a high degree of coherence for much longer than 10 fs \cite{GoulielmakisNature10}. Measuring the induced wave packet dynamics requires sub-femtosecond timing synchronization between pump and probe pulses, which has been experimentally demonstrated in attosecond science \cite{GoulielmakisNature10, BenedickNature12}. Here, we consider a coherent wave packet launched in a Br$_2$ molecule by a photoionizing pump pulse and imaged by an ultrafast x-ray probe pulse. 




  

A molecule or a solid studied by RXS should consist of rather heavy elements, which have a high $K$ (or $L$) edge energy, in order to gain a sufficiently high spatial resolution. For instance, resolutions better than 2.5 \r A may be obtained by pulses resonant with $K$ edges of elements with the atomic number $Z \ge 22$  or with $L$ edges of elements with $Z\ge 53$ \cite{BindingEnergies}. Therefore, we apply our study to electron hole dynamics in Br$_2$, which allows reaching a resolution of 0.9 \r A at the $K$ edge of Br.

\begin{figure}[t]
\includegraphics[width=0.4\textwidth]{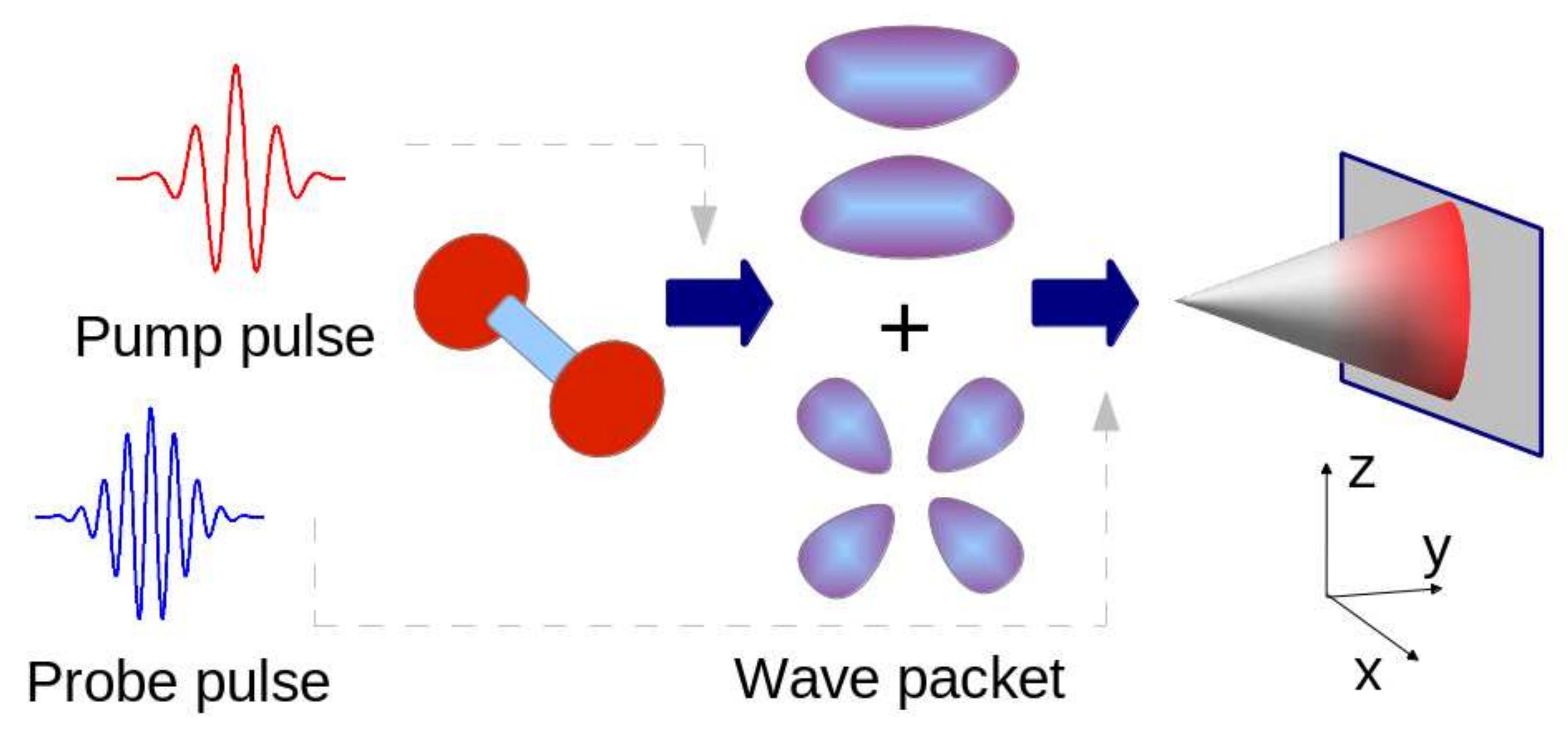}
  \caption{Schematic representation of the pump-probe scenario considered. The pump pulse excites a coherent superposition of the hole bonding and anti-bonding states of Br$_2$. A resonant x-ray scattering pattern is taken at time $t_p$ by the probe pulse.}
\label{Experiment}
\end{figure}

We treat Br$_2$ within the molecular orbital theory, the LCAO approximation, and  under the assumption that each atom contributes one atomic orbital to form a bond. Thus, the molecular bonding and antibonding orbitals of Br$_2$ can be expressed as \cite{Ballhausen}
\begin{align}
&|\phi_{\pm}\ra = \lf(|\widetilde\phi_a\ra\pm|\widetilde\phi_b\ra\rt)/\sqrt{2(1\pm S)},\label{HOMO}
\end{align}
with the corresponding energies $E_+$ and $E_{-}$. Here, $|\widetilde\phi_a\ra$ and $|\widetilde\phi_b\ra$ are basis wave functions localized at the two atoms of Br$_2$ denoted as $A$ and $B$, respectively; $S=\la\widetilde \phi_a|\widetilde\phi_b\ra$. The two highest occupied molecular orbitals of Br$_2$ are the $\pi_g$ and $\pi_u$ orbitals of Br $4p$ character (see Fig.~\ref{Fig_Br}a) \cite{PottsTrFSoc71}. Let us assume that the pump pulse created a hole initially localized at the $4p_z$ orbital of atom $A$ of Br$_2$ aligned in the $x$ direction (see Fig.~\ref{Experiment}). Such electron-hole localization in a molecule by a photoionizing pump pulse is possible as has been demonstrated in Ref.~\cite{SansoneNature10}. Then, the hole starts oscillating between sites $A$ and $B$ and at time $t_p$ is a superposition of the states (\ref{HOMO})
\begin{align}
|\Psi(t_p)\ra=C_+e^{-iE_+t_{\text{p}}}|\phi_{+}\ra+C_-e^{-iE_{-}t_{\text{p}}}|\phi_{-}\ra,\label{Wave_packet}
\end{align}
where $C_\pm = \sqrt{(1\pm S)/2}$. Strong-field ionization is highly selective in the sense that only the most weakly bound orbitals can be ionized \cite{PopovPhUsp04}. Therefore, orbitals that are more strongly bound than $\phi_+$ and $\phi_-$ can be treated as unaffected by ionization. The incident photon energy of the $z$-polarized probe pulse propagating in the $y$ direction is tuned close to the $K$ edge of Br: $\omega_\i\approx 13.5$ keV. Thereby, intermediate states are excited with a hole in either of the two Br $1s$ core orbitals, which we assume to be degenerate. Thus, assuming that the resonant excitation fills the valence hole, only two intermediate states play a role, associated with the two Br $1s$ orbitals $|\phi_{J_A}\ra$ and $|\phi_{J_B}\ra$ localized at $A$ and $B$, respectively (see Fig.~\ref{Fig_Br}b).

According to Eq.~(\ref{DSP_DipApp}), the DSP from wave packet (\ref{Wave_packet}) at time $t_p$ is 
\begin{align}
&\frac{dP}{d\Omega}= \frac{I_0\tau_p^2|D_{0z}|^2}{8\ln2c^4}\sum_{F,s_\s}\bigl\{|\mathbf D_{FJ_A}\cdot\boldsymbol\epsilon_\s^*|^2+|\mathbf D_{FJ_B}\cdot\boldsymbol\epsilon_\s^*|^2\label{DSP_DM}\\
&+2\Re[(\mathbf D_{FJ_A}\cdot\boldsymbol\epsilon_\s^*)(\mathbf D_{FJ_B}\cdot\boldsymbol\epsilon_\s^*)]\sin(Q_xR_x)\sin(2\pi t_p/T)\}\nonumber\\
&\quad\times\Delta\omega_{JF}^2\int_0^{\infty} \frac{d\omega_{\k_\s}\omega_{\k_\s}W(\omega_{\k_\s})e^{-\Omega_F^2\tau_p^2/4\ln2}}{\Gamma_J^2/4+(\omega_{\k_\s}-\Delta\omega_{JF})^2},\nonumber
\end{align}
where $R_x$ is the interatomic distance, $T=2\pi/(E_+-E_-)$, $D_{0z}=\int d^3r\widetilde\phi_a^*\,z\,\phi_{J_A}=\int d^3r\,\widetilde\phi_b^*\, z\,\phi_{J_B}$, $\mathbf D_{FJ_{A}(FJ_B)}=\int d^3r\phi_{J_{A}(J_B)}^*\mathbf r\,\phi_F$, 
$|\phi_F\ra$ is a molecular orbital where the hole is situated in the final state, $\Gamma_J=\Gamma_{J_{A}F}=\Gamma_{J_{B}F}$ and $\Delta\omega_{JF} = \Delta\omega_{J_{A}F}=\Delta\omega_{J_{B}F}$. There are two types of contributions to the scattering signal. The first two, time-independent terms in the curly braces in Eq.~(\ref{DSP_DM}) result in a constant background signal. The last term depends on the phase of the wave packet, $2\pi t_p/T$, and provides a $Q_x$-dependent diffraction signal. It determines the imaginary part of the Fourier transform of Eq.~(\ref{DSP_DM}) from $\mathbf Q$ space to real space, while the first two terms determine the real part. The time-dependent term is nonzero for transitions into final states for which both $\mathbf D_{FJ_A}$ and $\mathbf D_{FJ_B}$ are nonzero, {\it i.e.\!} into nonlocalizable states. Thus, both elastic and inelastic scattering processes involving final-state holes in outer shells contribute to the time-dependent term providing information about the wave packet.

\subsection{Experimental considerations}

\label{SectionExperiment}
\begin{table}[t]
\begin{center}
\begin{tabular}{| c | c |}
\hline
Process & Rate (atomic units)\\ 
\hline
$2p$ - $1s$ & $5.2 \times 10^{-2}$\\ 
$3p$ - $1s$ & $7.3 \times 10^{-3}$\\ 
$4p$ - $1s$ & $6.4 \times 10^{-4}$\\ 
\hline
\end{tabular}
\end{center}
\caption{Spontaneous emission rates in core-excited Br calculated using the XATOM toolkit \cite{SonPRA11}.}
\label{Emission_rates}
\end{table}

The background signal is dominated by inner-shell x-ray emission from the $2p\rightarrow 1s$ and $3p\rightarrow 1s$ transitions  contributing only to the time-independent terms. It follows from Table \ref{Emission_rates}, which shows the spontaneous emission rates in core-excited Br, that the ratio of non-signal to signal scattered photons is about 100:1. Since the photon energies due to inner-shell x-ray emission are much lower than the incoming photon energy, the detector window function $W(\omega_{\k_\s})$ should be centered at the incoming energy, $\omega_\i$, and be sufficiently narrow to suppress all $\omega_{\k_\s}\ll\omega_\i$ in order to decrease the background.

\begin{figure}[t]
\includegraphics{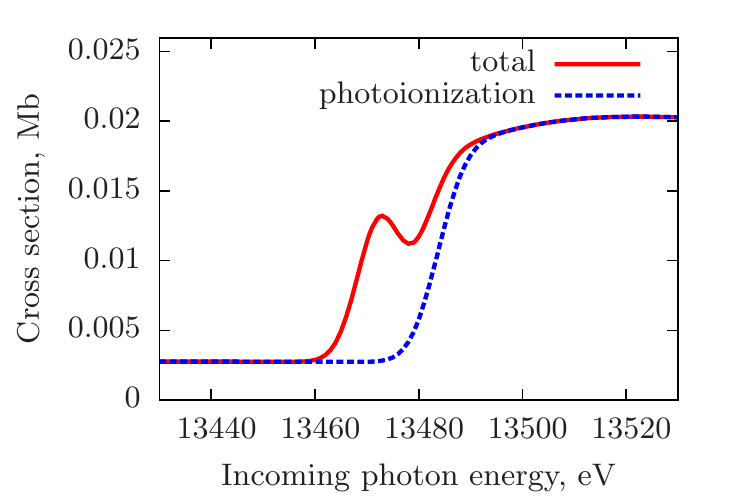}
  \caption{Photoionization cross section and total photoabsorption cross section of Br calculated using the XATOM toolkit \cite{SonPRA11}.}
\label{cross_sec}
\end{figure}

Another possible source of the background signal is due to excitation of core electrons into unoccupied states above the wave packet states. Therefore, the spectral width of the probe pulse must be small enough not to induce transitions into higher lying orbitals that do not contribute to the wave packet. But at the same time, the spectral width should be much larger than the energy splittings of the eigenstates comprising the wave packet, so that the probe-pulse duration is shorter than the characteristic time scale of dynamics in this wave packet. For our system, a probe pulse with a duration of 200 as still does not induce transitions into higher lying states of Br$_2$ and is much shorter than the period of the considered wave packet, $T \approx 2$ fs \cite{PottsTrFSoc71}. Scattering from molecules unaffected by the pump pulse and hence not ionized, does not contribute to the patterns, since there is no vacancy in the valence shell in this case so that resonant absorption is forbidden (see Fig.~\ref{Fig_Br}). Only a small fraction of absorbed photons lead to direct photoionization of orbitals above the $K$ shell. Fig.~\ref{cross_sec} shows the photoionization cross section and total photoabsorption (including photoionization and resonant photoexcitation) cross section for Br calculated using the XATOM toolkit \cite{SonPRA11}. It follows from Fig.~\ref{cross_sec} that, for photon energies near the Br $K$ edge, the photoionization cross section is negligible in comparison to the cross section for excitation of an electron from the $1s$ shell to the $4p$ shell in Br.

We now estimate the number of scattering patterns that is necessary to obtain a sufficiently high signal-to-noise ratio. Let us assume a peak x-ray pulse intensity $I_0=10^{18}$ W/cm$^2$ and pulse duration of 200 as corresponding to $10^{9}$ photons in a focal area $A_0$ of 1 $\mu$m$^2$. The solid angle corresponding to an independent pixel is
\begin{align}
d\Omega_p = \lf(\frac{\lambda}{R_x}\rt)^2=0.15,
\end{align} 
where $\lambda$ is the wavelength of the probe pulse and $R_x$ is the interatomic distance of Br$_2$. We find that $\la dP/d\Omega\ra\approx 2\times 10^{-8}$ [see Fig.~\ref{DSP_Fig} and the discussion of it in the next Subsection] and, thus, approximately $N_p = d\Omega_p\times\la dP/d\Omega\ra = 3\times 10^{-8}$ photons per pixel are scattered from a single molecule for each probe pulse. The preparation of an ensemble of aligned molecules provides a significant signal enhancement due to the possibility to average many identical patterns \cite{BartyARPC13}. An extremely high degree of molecular alignment can be achieved with a strong laser field (for instance, $\la \cos^2{\theta_{2D}}\ra=0.97$ in Ref.~\cite{HolmegaardPRL09} and $\la \cos^2{\theta_{2D}}\ra=0.89$ in Ref.~\cite{KupperPRL14}) \cite{StapelfeldtRMP03}. Alignment of Br$_2$ molecules has been discussed in Refs.~\cite{ButhPRA08,HoPRA08,HoJCP09} and demonstrated experimentally in Ref.~\cite{BertrandNature13}. Recently the feasibility of x-ray diffraction imaging of laser-aligned gas-phase molecules was demonstrated \cite{KupperPRL14}.

We assume a molecular beam density of $M = 10^{10}$ cm$^{-3}$, a molecular beam size of $l_m=0.4$ cm, an interaction area of $A_0= 10^{-8}$ cm$^2$ and 10\% of the molecules in the wave packet state, so that $M\times l_m\times A_0\times 10\% \approx 4$ molecules per shot contribute to a scattering pattern. Thus, roughly $10^7$ shots are necessary to obtain a signal of 1 photon per pixel. The necessary number of shots can be reduced to $10^5$ using the method in Ref.~\cite{FungNature08}, which allows reconstructing a structure of a single molecule with a mean photon count on the order of 0.01 photons per pixel. Thus, this experiment would, in principle, be feasible with the forthcoming European XFEL facility, which will provide 27 000 x-ray pulses per second \cite{BartyARPC13}. The only obstacle is that the European XFEL will initially not produce sub-femtosecond pulses, but this limitation can be overcome with the strategies described in Refs.~\cite{ZholentsPRL04, EmmaPRL04, SaldinPhysRevSTAB06, KumarAppSciences13, TanakaPRL13}.

\subsection{DSP from the oscillating electron hole in Br$_2$}

\begin{figure}[t]
\includegraphics[width=0.48\textwidth]{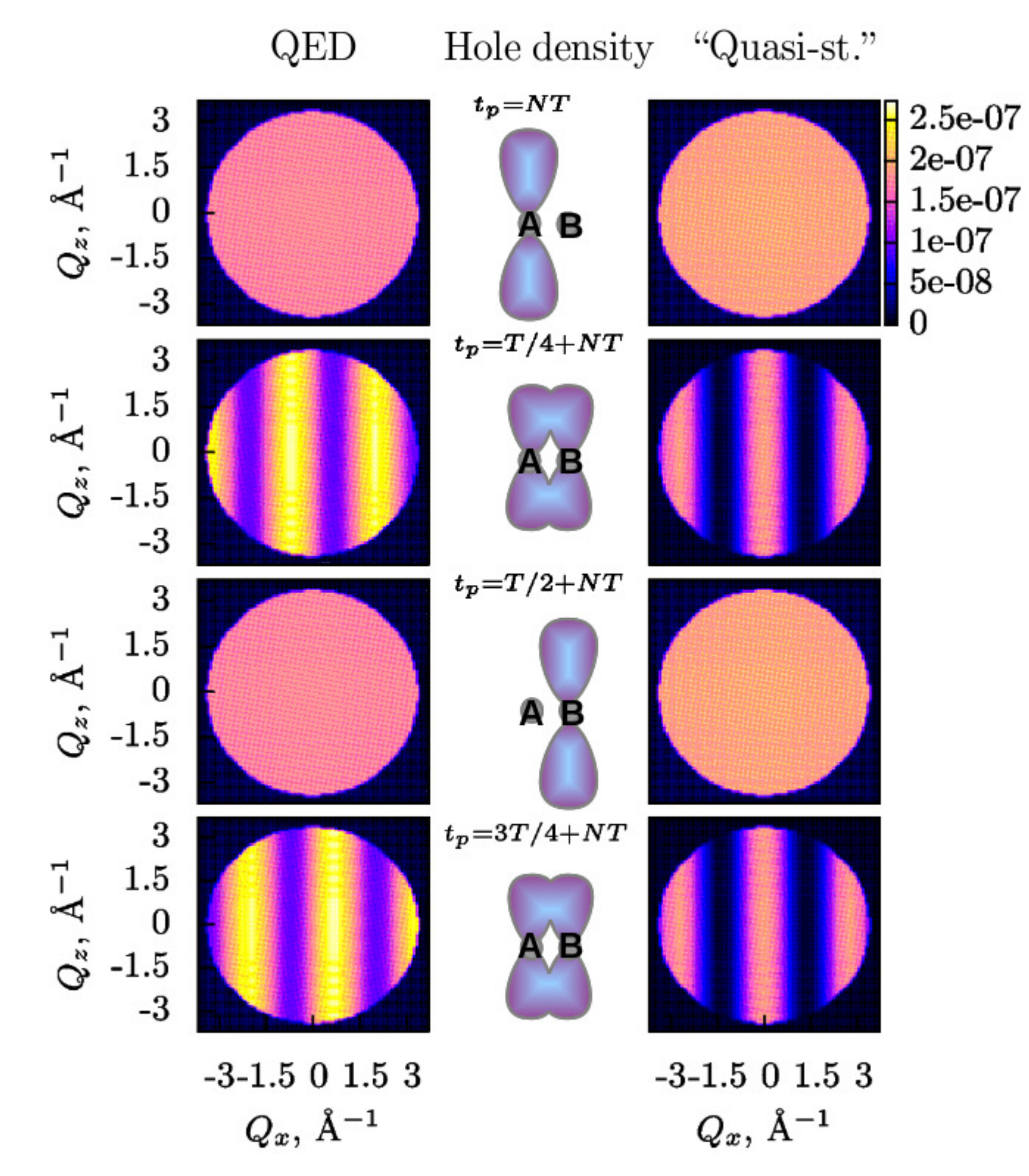}
\caption{DSP from a single molecule in the $Q_x-Q_z$ plane at $Q_y=0$  according to Eq.~(\ref{DSP_DM}) (left) and Eq.~(\ref{DSP_st}) (right) and the corresponding schematic representation of the hole density (middle) at different $t_p$. Br$_2$ parameters: Hole lifetime is $260\times10^{-18}$ s \cite{KrauseJPhChRD79}, $R_x=2.3$ \r A, $T=1.7$ fs \cite{PottsTrFSoc71}. Probe pulse parameters: $\tau_p=200$ as, $\omega_\i\approx13.5$ keV, $I_0=10^{18}$ W/cm$^2$. $N$ is a positive integer. The dependence on polarization is not shown. The ranges are limited by $Q_x^2+Q_z^2\leq \omega_\i^2/c^2$.}
\label{DSP_Fig}
\end{figure}

Figure~\ref{DSP_Fig} shows the DSP from the oscillating electron hole in Br$_2$ at different times $t_p$ according to Eqs.~(\ref{DSP_DM}) [QED description] and (\ref{DSP_st}) [``quasistationary" description]. For Eq.~(\ref{DSP_DM}), we assumed that $W(\omega_{\k_\s})$  is centered at $\omega_\i$ and its width is 5 eV, so that scattering only involving the $1s$ - $4p\pi_g$ and $1s$ - $4p\pi_u$ transitions is detected (see Fig.~\ref{Fig_Br}b). This assumption influences only the amplitude of the DSP, but not the basic features. The signals are constant when the hole is localized on one of the atoms (at times $0$ and $T/2$), since the scattering channel for the other atom is blocked at these moments, and there is no interference. A diffraction signal is obtained when the hole is delocalized, whereby the difference of the positions of the maxima is equal to $2\pi/R_x$ in both cases. The unidirectional structure of the DSP reflects that the molecule is aligned along the $x$ axis. Thus, at times $T/4$ and $3T/4$ in Figure~\ref{DSP_Fig}, the scattering patterns encode information about the alignment, orbital direction and interatomic distances of Br$_2$.

Surprisingly, the scattering patterns derived from QED depend on whether the hole is moving from site $A$ to site $B$ or vice versa. They are shifted relative to each other by $\pi /R_x$ and are not inversion-symmetric with respect to $\mathbf Q = 0$. Thus, a single scattering pattern provides the direction in which the electron hole is moving. In contrast, the DSP according to the ``quasistationary" description is the same at times $T/4$ and $3T/4$, when the instantaneous hole density is the same. This discrepancy demonstrates that the ``quasistationary" picture does not provide a correct description for a dynamical system. It fails because both elastic and inelastic processes are decisive for time-resolved measurements, for which the probe pulse bandwidth is much larger than energy splittings of electron states.

\begin{figure}[t]
\includegraphics{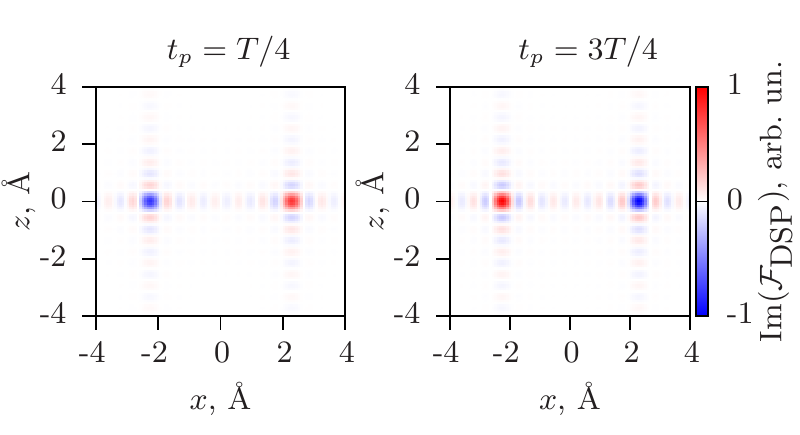}
\caption{Imaginary part of the spatial Fourier transform of the QED scattering patterns at times $T/4$ and $3T/4$.
 The interference fringes are due to the finite regions of the integrated functions.
 }
\label{Fur_Fig}
\end{figure}

The imaginary parts of the Fourier transforms of the QED scattering patterns are shown in Fig.~\ref{Fur_Fig} at times $t_p=T/4$ and $t_p=3T/4$. The amplitude and direction of the instantaneous interatomic electron current may be reconstructed by Fourier analysis of the scattering patterns. Let us calculate the imaginary part of Fourier transform of the DSP in Eq.~(\ref{DSP_DM}):
\begin{align}
\Im(\mathcal F_{\text{DSP}}) \propto &\sin(2\pi t_p/T)\lf(i\delta(x + R_x) - i\delta(x - R_x)\rt)\nonumber\\
&\times\sum_F\Re\bigl[(\mathbf D_{FJ_A}\cdot\boldsymbol\epsilon_\s^*)(\mathbf D_{FJ_B}\cdot\boldsymbol\epsilon_\s^*)\bigr].
\end{align}
Now, let us connect it to the electron hole current flowing from atom $A$ to atom $B$, $j_{ab}(t_p)$, and the electron hole current flowing  from atom $B$ to atom $A$, $j_{ba}(t_p)$,  derived in Appendix \ref{Appendix_InterCurrAB}:
\begin{align}
&\Im(\mathcal F_{\text{DSP}}) \propto j_{ab}(t_p)\delta(x-R_x)+j_{ba}(t_p)\delta(x+R_x)\\
&j_{ab}(t_p) = -j_{ba}(t_p) =\frac{\sin(2\pi t_p/T)}2 \int d^3 r \Re(\widetilde\phi_b^*\nabla_x\widetilde\phi_a)\nonumber.
\end{align}
Thus, the right delta peak appears at the position corresponding to the vector pointing from atom $A$ to atom $B$ and is proportional to the value of the electron hole current flowing  from atom $A$ to atom $B$, $j_{ab}(t_p)$ (see Eq.~(\ref{jab_App}) for details). The left peak is proportional to the value of the electron hole current flowing  from atom $B$ to atom $A$, $j_{ba}(t_p)$, which is opposite to $j_{ab}(t_p)$. At time $t_p=T/4$, when the electron hole is fully delocalized and flows from $A$ to $B$ [see Fig.~\ref{DSP_Fig}], the right peak reaches its maximum. The peaks are switched at time $t_p=3T/4$, when the electron hole flows in the opposite direction. The amplitudes of the peaks are proportional to the amplitudes of the currents. They decrease when the hole localizes on one of the atoms and are zero at times $t_p=0$ and $t_p=T/2$.

\section{Conclusions}

We have demonstrated that inelastic scattering, which unavoidably contributes to a scattering pattern obtained by an ultrashort probe pulse, can be used to one's advantage for imaging by time-resolved RXS. Applying a thorough theoretical analysis based on QED, we obtained that time-resolved scattering patterns are not determined by, but still connected to the instantaneous electron density and contain additional information about electron dynamics, resolving the direction of the electron flow. Thereby, due to the contribution of inelastic processes, information resolved in a scattering pattern in a time-resolved measurement differs from the information obtained under stationary conditions. This result is quite counterintuitive, since one may expect that a single snapshot from a dynamical system contains exactly the same amount of data as that from a stationary system. Our study demonstrates that such an interpretation is not correct, and when interpreting results obtained by time-resolved RXS, one should take into account that  scattering patterns are affected by inelastic processes and do not coincide with those obtained in a stationary measurement. We have shown that time-resolved RXS provides additional information about the dynamical properties of a nonstationary electronic system along with the instantaneous geometry, interatomic distances and directions of orbitals or spins within a sample. We have demonstrated how the direction of electron current can be retrieved from a scattering pattern obtained by RXS using Fourier analysis. 

Preparing and watching coherent electron wave packets in complex matter is an emerging opportunity in attosecond science. Current probes in attosecond science provide access to the time domain, but do not provide direct access to the spatial domain. We expect that time-resolved RXS will prove to be a powerful tool for time-resolved imaging providing access to the electronic motion that is the key to understanding various phenomena in complex molecular, biological and solid-state systems.


\section*{Acknowledgment}

We gratefully acknowledge the help of Sang-Kil Son in preparing Section \ref{SectionExperiment}.

\begin{appendix}
\section{Details of the QED derivation of the DSP}
\label{Appendix_DSP}

We study the resonant x-ray scattering from an electron system, which is described by the density matrix  
\begin{equation}
\hat \rho^{\text{m}}(t_0) = \sum_{I,K} \mathcal I_{IK}(t_0) |\Phi_I\ra\la \Phi_K|
\end{equation} 
at time $t_0$ after the action of a pump pulse. A scattering process consists in the annihilation of one photon and in the creation of one photon. Thus, the lowest-order contributing process demands that the operator $\hat{\mathbf A}$, which is linear in the creation and annihilation operators, acts twice. Therefore, we evaluate the total density operator within the second-order time-dependent perturbation theory using $\hat H_{\text{int}}$ as the perturbation:
\begin{align}
\hat \rho^{(2,2)} =& \sum_{\{n\},\{\widetilde n\},I,K}\rho ^X_{\{n\},\{\widetilde n\}}\mathcal I_{IK}(t_0)|\Psi_{\{n\}I}^{(2)},t_f\ra\la\Psi_{\{\widetilde n\}K}^{(2)},t_f|\label{rho22_App}.
\end{align}
Here, $\{n\}$ and $\{\widetilde n\}$ are complete sets that specify the number of photons in all initially occupied field modes with a distribution $\rho ^X_{\{n\},\{\widetilde n\}}$. $|\Psi_{\{n\}I}^{(2)},t_f\ra $ is the second-order wave function at time $t_f$, which is an entangled state of the electronic and photonic states:
\begin{align}
|\Psi_{\{n\}I}^{(2)},t_f\ra=&-\int_{t_0}^{t_f}dt'\,e^{i\hat H_0 t'}\hat H_{\text{int}}\,e^{-i\hat H_0 t'}\label{Psi2_App}\\
&\times\int_{t_0}^{t'}dt''\,e^{i\hat H_0 t''}\hat H_{\text{int}}\,e^{-i\hat H_0 t''}|\{n\}\ra|\Phi_I\ra,\nonumber
\end{align}
where $\hat H_0=H_{\text{m}}+\sum_{\mathbf{k},s}\omega_{\k,s}\hat a_{\k,s}^\dagger\hat a_{\k,s}$. $t_0$ and $t_f$ are times prior to and after the action of the probe pulse, thus, $t_0\to-\infty$ and $t_f\to+\infty$.

Probability $P(\omega_{\k_\s})$ according to Eq.~(\ref{DSP}) is
\begin{align}
&P(\omega_{\k_\s}) =\sum_{s_\s=1}^2 W(\omega_{\k_\s})\sum_{\{n'\},F}\la\Phi_F;\{n'\}|\hat \rho^{(2,2)}|\Phi_F;\{n'\}\ra\\
&=\sum_{s_\s=1}^2W(\omega_{\k_\s})\sum_{\{n'\},F}\sum_{\{n\},\{\widetilde n\}}\sum_{\k_1,\k_2,s_1,s_2}\sum_{I,K}\label{P_ot_T_App}\\
&\quad\times\rho ^X_{\{n\},\{\widetilde n\}} \mathcal I_{IK}(t_0)\sqrt{\frac{2\pi}{V\omega_{\k_1}}}\sqrt{\frac{2\pi}{V\omega_{\k_2}}}\frac{2\pi}{V\omega_{\k_\s}}\nonumber\\
&\quad\times\int_{t_0}^{t_f}dt'_1\int_{t_0}^{t_f}dt'_2\int_{t_0}^{t'_1}dt_1''\int_{t_0}^{t'_2}dt''_2\nonumber\\
&\qquad\times e^{i\omega_{\mathbf k_s}t_1'}e^{-i\omega_{\mathbf k_1}t_1''}e^{-i\omega_{\mathbf k_s}t_2'}e^{i\omega_{\mathbf k_2}t_2''}\nonumber\\
&\qquad\times \la\Phi_F|e^{i\hat H_{\text{m}} t'_1}\hat T^\dagger_{\mathbf k_s, s_s} e^{-i\hat H_{\text{m}} (t'_1-t''_1)}  \hat T_{\mathbf k_1, s_1}e^{-i\hat H_{\text{m}}t_1''}|\Phi_I\ra\nonumber\\
&\qquad\times
\la\Phi_K|e^{i\hat H_{\text{m}}t_2''}\hat T^\dagger_{\mathbf k_2, s_2}  e^{i\hat H_{\text{m}} (t'_2-t''_2)} \hat T_{\mathbf k_s, s_s} e^{-i\hat H_{\text{m}} t'_2}|\Phi_F\ra\nonumber\\
&\qquad\times\la\{n'\}|\hat a^\dagger_{\k_\s,s_\s} \hat a_{\k_1,s_1}|\{n\}\ra \la \{\widetilde n\}|  \hat a^\dagger_{\k_2,s_2}\hat a_{\k_\s,s_\s}|\{ n' \}\ra,\nonumber
\end{align}
where 
\begin{equation}
\hat T_{\mathbf k,s} = \int d^3 r \hat \psi^\dagger(\mathbf r)e^{i\mathbf k\cdot\mathbf r}(\boldsymbol\epsilon_{\mathbf k, s}\cdot \mathbf p)\hat \psi(\mathbf r)\label{TranOp_App}
\end{equation}
are transition operators. Below, we apply that 
\begin{align}
&\sum_{\{n\},\{\widetilde n\}}\rho ^X_{\{n\},\{\widetilde n\}} \sum_{\{n'\}}\la\{n'\}|\hat a^\dagger_{\k_\s,s_\s} \hat a_{\k_1,s_1}|\{n\}\ra \label{Trace_a12_App}\\
&\quad\times\la \{\widetilde n\}|  \hat a^\dagger_{\k_2,s_2}\hat a_{\k_\s,s_\s}|\{ n' \}\ra=\Tr[\hat \rho_{\text{in}}^X\hat a^\dagger_{\mathbf k_2,s_2}\hat a_{\mathbf k_1,s_1}]\nonumber,
\end{align}
where $\hat\rho_{\text{in}}^X=\sum_{\{n\},\{\widetilde n\}} \rho_{\{n\},\{\widetilde n\}}^X  |\{n\}\ra\la\{\widetilde n\}|$ is the initial density operator of the radiation field \cite{Loudon, Mandel}.

We assume that the bandwidth and the angular spread of the x-ray pulse are sufficiently small to satisfy the conditions $\sqrt {\omega_{\k_1}\omega_{\k_2}}\approx \omega_{\i}$, $\k_{1,2}\approx\k_\i$ and $\boldsymbol\epsilon_{\k_{1,2}}\approx\boldsymbol\epsilon_\i$, where  $\omega_{\i}$, $\k_\i$ and $\boldsymbol\epsilon_\i$ are the mean values of the photon energy, wave vector and polarization of the incident beam, respectively. Therefore,
\begin{align}
&\frac{2\pi\omega_{\text{in}}}{V}\sum_{\mathbf k_1,\mathbf k_2, s_1,s_2}\Tr[\hat \rho_{\text{in}}^X\hat a^\dagger_{\mathbf k_2,s_2}\hat a_{\mathbf k_1,s_1}]e^{-i\omega_{\mathbf k_1}t_1''}e^{i\omega_{\mathbf k_2}t_2''}\label{CorrFun_App}\\
&=\frac{2\pi}{c} I_0(\mathbf r_0)e^{-2\ln2\lf(\frac{t_1''-t_p}{\tau_p}\rt)^2}e^{-2\ln2\lf(\frac{t_2''-t_p}{\tau_p}\rt)^2}e^{i\omega_{\text{in}}(t_2''-t_1'')}\nonumber,
\end{align}
Here, the probe pulse has a Gaussian shape and the amplitude of the electric field is $E(\mathbf r_0,t)=\sqrt{(8\pi/c) I_0(\mathbf r_0)}e^{-2\ln2\lf(\frac{t-t_p}{\tau_p}\rt)^2}$, $\mathbf r_0$ is the position of the object, $t_p$ is the time of the measurement, $\tau_p$ is the pulse duration (FWHM of the pulse intensity) and $I_0(\mathbf r_0)=cE^2(\mathbf r_0,t=0)/(8\pi)$. Note that expression (\ref{CorrFun_App}) is the first-order radiation field correlation function \cite{GlauberPhRev63, Loudon}. Since the probe pulse must be much shorter than the time variation of the electron density during the action of the probe pulse, the following condition must be satisfied for an appropriate time-resolved measurement
\begin{align}
&e^{-i\hat H_{\text{m}}t_1''}\hat \rho^{\text{m}}(t_0)e^{i\hat H_{\text{m}}t_2''} \approx e^{i\la E \ra (t_2''-t_1'')}\hat \rho^{\text{m}}(t_p),\label{UltrashortPulseApr_App}
\end{align}
where $\hat \rho^{\text{m}}(t_p) = \sum_{I,K} \mathcal I_{IK}(t_p)|\Phi_I\ra\la \Phi_K|$ is the electron density at the time of the measurement and $\la E \ra$ is the mean energy of the nonstationary electron system. Therefore, the DSP at the time of measurement $t_p$ is
%
\begin{align}
\frac{dP}{d\Omega}=&\frac{I_0}{2\pi\omega_{\i}^2c^4}\sum_{s_\s=1}^2\int_0^{\infty}d\omega_{\k_\s}\omega_{\k_\s}W(\omega_{\k_\s})\label{DSP_final_App}\\
&\times\sum_F\sum_{I,K}\mathcal I_{IK}(t_p)\la \Phi_K| \hat{\mathcal T}^\dagger|\Phi_F\ra\la\Phi_F| \hat{\mathcal T} |\Phi_I\ra\nonumber\\
\hat{\mathcal T} = &\int_{t_0}^{t_f}dt'\int_{t_0}^{t'}dt'' e^{-2\ln2\lf(\frac{t''-t_p}{\tau_p}\rt)^2}e^{-i\omega_{\text{in}}t''}e^{i\omega_{\mathbf \k_\s}t'}\nonumber\\
&\times e^{i\hat H_{\text{m}} t'}\hat T^\dagger_{\mathbf k_s, s_s} e^{-i\hat H_{\text{m}} (t'-t'')}  \hat T_{\mathbf k_\i, s_\i}e^{-i\la E \ra t''}\nonumber.
\end{align}
Note that $|\Phi_F\ra$, $|\Phi_I\ra$ and $|\Phi_K\ra$ can be either the same or different eigenstates of $\hat H_{\text{m}}$.

\subsection{Electric dipole approximation}

We consider a resonant process involving transitions of core electrons to valence shells. Core shells are highly localized in comparison to the resonant wavelengths considered here. The size of a core shell scales with the effective nuclear charge $Z_{\text{eff}}$ ($Z_{\text{eff}}$ for K-shells is approximately equal to the nuclear charge) as $n^2a_0/Z_{\text{eff}}$, where  $a_0=0.529$ \r A is the Bohr radius, $n$ is the number of the shell \cite{LandauQuantum}. Therefore, we consider the spatial distribution of the electric field, $e^{i\k\cdot\mathbf r}$,  within the dipole approximation for each absorbing atom. Thus, the term $e^{i\k\cdot\mathbf r}$ may be approximated by $e^{i\k\cdot\mathbf R}$, where $\mathbf R$ is the position of the core shell, from which the electron is excited. This leads to
\begin{align}
&\la\Phi_F|e^{i\hat H_{\text{m}} t'}\hat T^\dagger_{\mathbf k_\s, s_\s} e^{-i\hat H_{\text{m}} (t'-t'')} \hat T_{\mathbf k_\i, s_\i}e^{-i\la E \ra t''}|\Phi_I\ra\nonumber\\
&=-\sum_{J_C}e^{iE_F t'}e^{i(E_{J_C}-i\Gamma_{J_C}/2)(t''-t')}e^{-i\la E \ra t''}
\\%
&\quad\times\la \Phi_F|\hat \psi^\dagger  e^{-i\mathbf k_{\s}\cdot\mathbf R_{C}}(\boldsymbol\epsilon_{\s}^*\cdot \mathbf p)\hat \psi |\Phi_{J_C}\ra
\nonumber\\
&\quad\times\la\Phi_{J_C}|\hat \psi^\dagger e^{i\mathbf k_{\i}\cdot\mathbf R_{C}} (\boldsymbol\epsilon_{\i}\cdot \mathbf p)\hat \psi|\Phi_I\ra
\nonumber
\end{align}
where $|\Phi_{J_C}\ra$ is an intermediate state with a hole in a core shell localized at $\mathbf R_{C}$. Since a core excited state exists for a very short time $\tau_{J_C}$, its line width $\Gamma_{J_C}=1/\tau_{J_C}$ is taken into account. 
After the integration over time and with the relation $\mathbf p = -i[\mathbf r, \hat H_{\text{m}}]$, the DSP within the dipole approximation becomes
\begin{align}
&\frac{dP}{d\Omega}= \frac{\tau_p^2I_0}{4\ln2c^4}\sum_{I,K}\mathcal I_{IK}(t_p)\label{DSP_DipApp_App}\\
&\qquad\times\sum_{C_q,C_r}e^{i\mathbf Q\cdot(\mathbf R_{C_q}-\mathbf R_{C_r})}\la \Phi_K |  \hat {\mathcal G}_{qr} |\Phi_I\ra,\nonumber\\ 
&\hat {\mathcal G}_{qr} = \hat d_{C_r}^\dagger \hat S  \hat d_{C_q},\label{Gqr_App}\\
&\hat d_{C} =\sum_{J_C}|\Phi_{J_C}\ra\la\Phi_{J_C}| \boldsymbol\epsilon_\i\cdot\hat{\mathbf r},\nonumber\\
&\hat S = \sum_{s_\s=1}^2\int_0^{\infty}d\omega_{\k_\s}\omega_{\k_\s}W(\omega_{\k_\s})\sum_F\hat {\mathcal L}^{\dagger}|\Phi_F\ra\la\Phi_F|\hat {\mathcal L},\nonumber\\
&\la\Phi_F|\hat {\mathcal L}|\Phi_{J_C}\ra = \frac{e^{-\frac{\Omega_F^2\tau_p^2}{8\ln2}}\Delta\omega_{J_{C}F}\la\Phi_F|\boldsymbol\epsilon_\s^*\cdot\hat{\mathbf r}|\Phi_{J_C}\ra}
{((\omega_{\k_\s}-\Delta\omega_{J_{C}F})+i\Gamma_{J_{C}}/2)}\nonumber
,
\end{align}
where $\mathbf Q=\k_{\i}-\k_{\s}$ is the scattering vector, $\Omega_F=\omega_{\k_\s}-\omega_{\i}+E_F-\la E \ra $, $\Delta\omega_{J_CF}=E_{J_C}-E_F$ and $\hat{\mathbf r}  = \int d^3 r \psi^\dagger(\mathbf r)\,\mathbf r\,\psi(\mathbf r)$.
Here, we applied that the transition energies by the absorption process can be approximated by $\omega_{\i}$ because of the resonance condition.

If the system is in a coherent wave packet state, when the matrix elements of the instantaneous electron density can be represented as $\mathcal I_{IK}(t_p)=C_IC_K^*e^{-i(E_I-E_K)t_p}$, and the Eq.~(\ref{DSP_DipApp_App}) reduces to
\begin{align}
&\frac{dP}{d\Omega}= \frac{ I_0\tau_p^2}{4\ln2c^4}\int_0^{\infty}d\omega_{\k_\s}\omega_{\k_\s}W(\omega_{\k_\s})\sum_{F,s_\s}\Biggl|\sum_{J_C}\Delta\omega_{J_CF}
\label{DSP_DipAppPure_App}\\
&\times\frac{\la \Phi_F|\boldsymbol\epsilon_\s^*\cdot\hat{\mathbf r}|\Phi_{J_C}\ra\la\Phi_{J_C}|\boldsymbol\epsilon_\i\cdot\hat{\mathbf r}|\Psi(t_p)\ra }{(\omega_{\k_\s}-\Delta\omega_{J_CF})+i\Gamma_{J_C}/2} e^{i\mathbf Q\cdot\mathbf R_{C}}\Biggr|^2 e^{-\frac{\Omega_F^2\tau_p^2}{4\ln2}},
\nonumber
\end{align}
where $|\Psi(t_p)\ra = \sum_I C_I e^{-iE_It_p}|\Phi_I\ra$.

\subsection{Application to a stationary system}
\label{App_Stationary}

In this Subsection, we show that Eq.~(\ref{DSP_DipApp}) converts into the conventional equation of the DSP for stationary systems. For a stationary system in the state $G$, $|\Psi(t_p)\ra$ is substituted by $|\Phi_G\ra$ in Eq.~(\ref{DSP_DipAppPure_App}), which leads to
\begin{align}
&\frac{dP}{d\Omega}= 
\frac{\int dt I_\i(t)}{\omega_\i c^4}\sum_{s_\s=1}^2\int_0^{\infty}d\omega_{\k_\s}\omega_{\k_\s} \omega_\i W(\omega_{\k_\s})\nonumber\\
&\times\sum_F\frac{\tau_pe^{-\frac{(\omega_{\k_\s}-\omega_\i+E_F-E_G)^2\tau_p^2}{4\ln2}}}{2\sqrt{\pi\ln2}}\label{DSP_st_short_App}
\\
&\times\Biggl|\sum_{J_C}
\frac{\Delta\omega_{J_CF}\la \Phi_F| \boldsymbol\epsilon_\s^*  \cdot\hat {\mathbf r}|\Phi_{J_C}\ra\la\Phi_{J_C}| \boldsymbol\epsilon_\i\cdot\hat {\mathbf r}|\Phi_G\ra }{(\omega_{\k_\s}-\Delta\omega_{J_CF})+i\Gamma_{J_C}/2}e^{i\mathbf Q\cdot\mathbf R_{C}}
 \Biggr|^2\nonumber.
\end{align}
Here, we expressed $I_0$ via the integral $\int dt I_\i(t)$, where $I_\i(t)=I_0e^{-4\ln2(t-\tau_p)^2/t^2}$ is the intensity of the incoming beam. 

The term $\tau_pe^{-\frac{(\omega_{\k_\s}-\omega_\i+E_F-E_G)^2\tau_p^2}{4\ln2}}/(2\sqrt{\pi\ln2})$ becomes the Dirac delta function at $\tau_p\gg\tau_{J_C}$. Thus, Eq.~(\ref{DSP_DipApp_App}) for a stationary system and at $\tau_p\gg\tau_{J_C}$ becomes
\begin{align}
\frac{dP}{d\Omega}&= 
\frac{\int dt I_\i(t)}{\omega_\i }\sum_{s_\s=1}^2{\frac{d\sigma}{d\Omega}}^{\text{st}},
\end{align}
where
\begin{align}
&{\frac{d\sigma}{d\Omega}}^{\text{st}}=\frac{\omega_\i}{c^4}
\sum_{F}\Biggl|\sum_{J_C} \frac{\la \Phi_F| \boldsymbol\epsilon_\s^*  \cdot\hat {\mathbf r}|\Phi_{J_C}\ra\la\Phi_{J_C}| \boldsymbol\epsilon_\i\cdot\hat {\mathbf r}|\Phi_G\ra }{(\omega_{\k_\s}-\Delta\omega_{J_CF})+i\Gamma_{J_C}/2}\\
&\times e^{i\mathbf Q\cdot\mathbf R_{C}}\Delta\omega_{J_CF}
 \Biggr|^2 W(\omega_{\k_\s})\omega_{\k_\s}\delta(\omega_{\k_\s}-\omega_\i+E_F-E_G)\nonumber
\end{align}
is the conventional expression for the stationary differential scattering cross section by resonant x-ray scattering of long pulses ($\tau_p\gg\tau_{J_C}$) taking into account both elastic ($F=G$) and inelastic ($F\neq G$) contributions \cite{MaPRB94}. Note that this relation is not correct for a stationary system, if $\tau_p\sim\tau_{J_C}$, and one should apply Eq.~(\ref{DSP_st_short_App}) in this case.

\section{Probability current density}
\label{App_Current}

The field annihilation (creation) operators $\psi(\mathbf r)(\psi^{\dagger}(\mathbf r))$ can be expanded in terms of one-particle wave functions $\phi_\alpha(\mathbf r)$:
\begin{align}
&\hat\psi(\mathbf r) = \sum_{\alpha}\hat c_{\alpha}\phi_\alpha(\mathbf r),\\
&\hat\psi^{\dagger}(\mathbf r) = \sum_{\alpha}\hat c_{\alpha}^{\dagger}\phi^{\dagger}_\alpha(\mathbf r),
\end{align}
where $\hat c_{\alpha}(\hat c_{\alpha}^{\dagger})$ annihilates (creates) a particle with a wave function $\phi_\alpha(\mathbf r)$. Thus, the probability current density in Eq.~(\ref{ProbCurrDen}) can be represented as
\begin{align}
\mathbf j(\mathbf r, t_p)  =\Im\lf(\sum_{I,K}\mathcal I_{IK}\sum_{\alpha,\beta}\la\Phi_K|\hat c_\beta^\dagger\phi_{\beta}^*(\mathbf r)\boldsymbol\nabla\hat c_\alpha\phi_\alpha(\mathbf r)|\Phi_I\ra \rt).
\end{align}
If one-particle wave function $\phi_{\alpha}$ can be represented as a linear combination of functions $\widetilde \phi_i(\mathbf r - \mathbf R_C)$ centered at site $ \mathbf R_C$,
\begin{align}
&\phi_{\alpha}(\mathbf r) = \sum_C\sum_i\gamma_{\alpha, Ci}\widetilde\phi_i(\mathbf r - \mathbf R_C)\label{onepart_app},
\end{align}
then $\mathbf j(\mathbf r, t_p)$ can be expressed as
\begin{align}
\mathbf j(\mathbf r, t_p) =&\sum_{C_a,C_b}\Im\lf(\sum_{I,K}\mathcal I_{IK}\la\Phi_K|\hat\xi_{C_b}^\dagger(\mathbf r)\boldsymbol\nabla\hat\xi_{C_a}(\mathbf r)|\Phi_I\ra\rt)\label{Curr_sumat_App},
\end{align}
where the operator
\begin{align}
&\hat \xi_{C}(\mathbf r) = \sum_{\alpha}\sum_i\gamma_{\alpha, Ci}\widetilde\phi_i(\mathbf r - \mathbf R_{C})\hat c_{\alpha}
\end{align}
annihilates an electron localized at atom $C$. Eq.~(\ref{Curr_sumat_App}) can be decomposed into intraatomic ($C_a=C_b$) and interatomic ($C_a\neq C_b$) contribution. Then, the volume averaged electron current between scattering atoms $C_q$ and $C_r$ is
\begin{align}
&j_{qr}(t_p) = \Im\Bigl (\sum_{I,K}\mathcal I_{IK}\sum_{\alpha,\beta}\la\Phi_K|\hat c_{\beta}^\dagger\hat c_{\alpha}|\Phi_I\ra\sum_{i,k}\gamma_{\alpha, C_q i }\gamma_{\beta, C_r k}^*\nonumber\\
&\quad\times \int d^3 r\widetilde\phi_k^* (\mathbf r - \mathbf R_{C_r})(\boldsymbol\nabla\cdot\mathbf n_{qr}) \widetilde \phi_i(\mathbf r - \mathbf R_{C_q}) \Bigr)\label{intercurr_App} ,
\end{align}
where $\mathbf n_{qr}$ is a unit vector pointing from site $C_q$ to site $C_r$.

\subsection{Connection to a factor $\mathcal J_{qr}(t_p)$}
\label{Connection_SecApp}

According to Eq.~(\ref{DSP_DipApp_App}), a factor $\mathcal J_{qr}(t_p) = \Im\lf(\sum_{I,K}\mathcal I_{IK}(t_p)\la \Phi_K|\mathcal {\hat G}_{qr}|\Phi_I\ra\rt)$ can be represented as
\begin{align}
\mathcal J_{qr}(t_p) = &H_{qr} \Im\Biggl (\sum_{I,K}\mathcal I_{IK}(t_p)\\
&\quad\times\la\Phi_K|\boldsymbol\epsilon_\i^*\cdot \hat {\mathbf r}|\Phi_{J_{C_r}}\ra   \la \Phi_{J_{C_q}}|\boldsymbol\epsilon_\i\cdot \hat {\mathbf r}|\Phi_I\ra\Biggr),\nonumber
\end{align}
where the term
\begin{align}
H_{qr} = &\sum_F\Delta\omega_{J_{C}F}^2\la\Phi_{J_{C_r}}|\boldsymbol\epsilon_\s\cdot \hat {\mathbf r}|\Phi_F\ra\la\Phi_F|\boldsymbol\epsilon_\s^*\cdot \hat {\mathbf r}|\Phi_{J_{C_q}}\ra\nonumber\\
&\quad\times\int_0^{\infty} \frac{d\omega_{\k_\s}\omega_{\k_\s}W(\omega_{\k_\s})e^{-\frac{\Omega_F^2\tau_p^2}{4\ln2}}}
{(\omega_{\k_\s}-\Delta\omega_{JF})^2+\Gamma_{J}^2/4}
\end{align}
does not depend on states $I$ and $K$, and, therefore, does not influence the temporal behavior of $\mathcal J_{qr}(t_p)$. Here, we applied that the x-ray probe pulse is resonant with a transition energy corresponding to the excitation of a core shell electron, and therefore, the contributions of other transitions is negligible. We also assume that states $J_{C_q}$ and $J_{C_r}$ are degenerate and differ only by the atom, at which the electron hole is localized (as a result, $\Delta \omega_{J_{C_q}F} = \Delta \omega_{J_{C_r}F} = \Delta \omega_{JF}$ and $\Gamma_{J_{C_q}}=\Gamma_{J_{C_r}}=\Gamma_{J}$), since the energy splittings of core-excited states would be much lower than the bandwidth of an ultrashort probe pulse.

Let us apply the representation of a one-particle wave function $\phi_{\alpha}$ given in Eq.~(\ref{onepart_app}) to $\mathcal J_{qr}(t_p)$:
\begin{align}
&\mathcal J_{qr}(t_p) = H_{qr}\sum_{\alpha,\alpha',\beta,\beta'}\Im\Biggl (\sum_{I,K}\mathcal I_{IK}\la\Phi_K|\hat c_{\beta}^\dagger \hat c_{\beta'} |\Phi_{J_{C_r}}\ra  \nonumber\\
&\times \la \Phi_{J_{C_q}}|\hat c_{\alpha'} ^\dagger    \hat c_{\alpha}|\Phi_I\ra\Biggr)\sum_{C_a,C_b,i,k}\gamma^*_{\beta,C_b k} \gamma_{\alpha,C_ai}d^*_{C_ak,C_r}d_{C_ai,C_q},\nonumber\\
&d^*_{C_bk,C_r}=  \int d^3 r\widetilde\phi_k^* (\mathbf r - \mathbf R_{C_b})(\boldsymbol\epsilon_\i^*\cdot\mathbf r) \widetilde\phi_{\text{core}}(\mathbf r - \mathbf R_{C_r}),\\
&d_{C_ai,C_q} = \int d^3r \widetilde\phi_i(\mathbf r - \mathbf R_{C_a})(\boldsymbol\epsilon_\i\cdot\mathbf r)\widetilde\phi^*_{\text{core}}(\mathbf r - \mathbf R_{C_q}).\nonumber
\end{align}
The action of the operator $\hat c_{\alpha'}^\dagger\hat c_\alpha$ leads to a creation of an electron hole in a core shell of atom $C_q$ with a wave function $\widetilde \phi_{\text{core}}$ centered at atom $C_q$ and annihilation of an electron hole in a valence shell by an electron with wave function $\sum_{C_a,i} \gamma_{\alpha C_a i}\widetilde \phi_{i}(\mathbf r - \mathbf R_{C_a})$ (analogously for $\hat c_{\beta}^\dagger\hat c_{\beta'}$). The integrals $d_{C_ai,C_q} $ for $C_a\neq C_q$ and $d_{C_bk,C_r} $ for $C_b\neq C_r$ are negligible in comparison to the integrals $d_{C_qi,C_q} $ and $d_{C_rk,C_r} $, therefore,
\begin{align}
\mathcal J_{qr}\approx &H_{qr}\sum_{\alpha,\alpha',\beta,\beta'}\Im\lf (\sum_{I,K}\mathcal I_{IK}\la\Phi_K|\hat c_{\beta}^\dagger \hat c_{\beta'} |\Phi_{J_{C_r}}\ra \rt.\label{Jqr_via_onepart_App}\\
&\times\lf.  \la \Phi_{J_{C_q}}|\hat c_{\alpha'} ^\dagger    \hat c_{\alpha}|\Phi_I\ra\rt)\sum_{i,k}\gamma^*_{\beta,C_r k} \gamma_{\alpha,C_qi}\nonumber\\
& \times\lf(\int d^3 r\widetilde\phi_k^* (\mathbf r - \mathbf R_{C_r})(\boldsymbol\epsilon_\i^*\cdot\mathbf r)  \widetilde\phi_{\text{core}}(\mathbf r - \mathbf R_{C_r})\rt)\nonumber\\
& \times\lf(\int d^3r \widetilde\phi_i(\mathbf r - \mathbf R_{C_q})(\boldsymbol\epsilon_\i\cdot\mathbf r)\widetilde\phi^*_{\text{core}}(\mathbf r - \mathbf R_{C_q})\rt)\nonumber.
\end{align}

The probe x-ray pulse must excite an electron from a core shell to electron states, where electron dynamics takes place. In this case, the terms in the sum over $\alpha,\beta$ in Eq.~(\ref{Jqr_via_onepart_App}) are nonzero for the same $\alpha$ and $\beta$ that enter interatomic electron hole current between atoms $C_q$ and $C_r$ (see Eq.~(\ref{intercurr_App})), and, thus, the factors $\gamma^*_{\beta,C_r k} \gamma_{\alpha,C_qi} $ entering $\mathcal J_{qr}$ and $j_{qr}$ would be the same. Hence, the factor $\mathcal J_{qr}$ would be exactly proportional to $j_{qr}$ at any $t_p$ under approximation that each atom contributes one atomic orbital to a molecular orbital, since there is only one term in the sum over $i$ and $k$. If it is not the other case,  $\mathcal J_{qr}$ would be proportional to $j_{qr}$, as long as for every $i$ and $k$, the ratio between
$d_{C_qi,C_q}d_{C_rk,C_r}^*$ and $ \int d^3 r\widetilde\phi_k^* (\mathbf r - \mathbf R_{C_r})(\boldsymbol\nabla\cdot\mathbf n_{qr}) \widetilde \phi_i(\mathbf r - \mathbf R_{C_q})$ is equal.

\subsection{Probability current density in Br$_2$}

\label{Appendix_InterCurrAB}
According to Eq.~(\ref{InterAtCurr}), the interatomic electron hole currents in the system described in Section \ref{DiatMolecule} are
\begin{align}
j_{ab} & = \Im\Biggl[  \int d^3 r\biggl( C_+^*C_-e^{i(E_+-E_-)t_p}\la \phi_+|\hat\xi_{B}^\dagger \nabla_x \hat\xi_{A} |\phi_-\ra \nonumber\\
&\qquad+C_+C_-^*e^{-i(E_+-E_-)t_p}\la \phi_-|\hat\xi_{B}^\dagger \nabla_x \hat\xi_{A} |\phi_+\ra\biggr)\Biggr]\nonumber\\
&=\frac12\sin(2\pi t_p/T)\int d^3 r \Re\lf( \widetilde\phi^*_b \nabla_x \phi_a \rt)\label{jab_App}\\
j_{ba} & = \frac12\sin(2\pi t_p/T)\int d^3 r \Re\lf( \widetilde\phi^*_a (-\nabla_x) \phi_b \rt) = -j_{ab},\nonumber
\end{align}
where $j_{ab}(t_p)$ is the electron hole current from atom $A$ to atom $B$, $j_{ba}(t_p)$ is the electron hole current from atom $B$ to atom $A$, and $T = 2\pi/(E_+-E_-)$. The signs of the gradients are opposite for $j_{ab}$ and $j_{ba}$, because $\mathbf n_{AB} = -\mathbf n_{BA}$.

\end{appendix}

\end{document}